\documentclass[
  twocolumn,
  prb,
  showpacs,
  amsmath,
  amssymb,
  superscriptaddress
]{revtex4}

\usepackage{bm}
\usepackage{graphicx}

\usepackage{subfigure}    

\newcommand{\Ci}{\mathop{\mathrm{Ci}}}
\newcommand{\Si}{\mathop{\mathrm{Si}}}

\renewcommand{\Re}{\mathop{\mathrm{Re}}}
\begin{document}

\title{
Measuring the Luttinger liquid parameter with shot noise
}

\author{J.\ K.\ K\"uhne}
\affiliation{
 Institut f\"ur Nanotechnologie, Karlsruhe Institute of Technology,
 76021 Karlsruhe, Germany
}
\affiliation{
 Institut f\"ur Theorie der Kondensierten Materie, 
Karlsruhe Institute of Technology, 76128 Karlsruhe, Germany}

\author{I.\ V.\ Protopopov}
\affiliation{
 Institut f\"ur Theorie der Kondensierten Materie, 
Karlsruhe Institute of Technology, 76128 Karlsruhe, Germany}
\affiliation{
 L.\ D.\ Landau Institute for Theoretical Physics RAS,
 119334 Moscow, Russia
}

\author{Y.\ Oreg}
\affiliation{
 Department of Condensed Matter Physics, Weizmann Institute of Science, Rehovot, Israel 76100
}

\author{A.\ D.\ Mirlin}

\affiliation{
 Institut f\"ur Nanotechnologie, Karlsruhe Institute of Technology,
 76021 Karlsruhe, Germany
}
\affiliation{
Institut f\"ur Theorie der Kondensierten Materie, 
Karlsruhe Institute of Technology, 76128 Karlsruhe, Germany}


\affiliation{
 Petersburg Nuclear Physics Institute,  188300 St.~Petersburg, Russia.
}

\begin{abstract}
We explore the low-frequency noise of interacting electrons in a one-dimensional structure (quantum wire or interaction-coupled edge states) with counterpropagating modes, assuming a single channel in each direction. The system is driven out of equilibrium by a quantum point contact (QPC) with an applied voltage, which induces a double-step energy distribution of incoming electrons on one side of the device. A second QPC serves to explore the statistics of outgoing electrons. We show that measurement of a low-frequency noise in such a setup allows one to extract  the Luttinger liquid constant $K$ which is the key parameter characterizing an interacting 1D system. We evaluate the dependence of the zero-frequency noise on $K$ and on parameters of both QPCs (transparencies and voltages). 
 
\end{abstract}


\maketitle
\section{Introduction}

The physics of interacting electrons in one dimension (1D) is profoundly different from  that in  higher dimensions. 
It is well known that  the correspondence between interacting electrons and free fermionic quasiparticles,
 which is in the core of  Landau's Fermi-liquid theory, breaks down in 1D. 
The resulting strongly correlated state, known as the  Luttinger liquid (LL),  can not be treated by conventional Fermi-liquid methods. Fortunately, there 
exists an extremely powerful approach to the problem, the bosonization technique\cite{Tomonaga,Luttinger,Giamarchi,Gogolin,Delft,Haldane}. It describes the low-energy sector of the theory in terms of density fluctuations, which are, under the simplest circumstances, non-interacting bosons. 

A key parameter invoked in  the bosonization description of a LL state is the interaction constant $K$. 
This dimensionless parameter gives an effective measure of the strength of the interaction between the electrons, 
with $K=1$ corresponding to a non-interacting Fermi gas, $K< 1$ to repulsion, and $K>1$ to attraction. 
The LL constant $K$ controls the behavior of various physical properties of the system \cite{Giamarchi}, 
including, e.g., the scaling  of the tunneling density of states  away from the wire ends (TDOS)\cite{LutherPeschel1974}, $\nu(\epsilon)\propto |\epsilon|^{(1-K)^2/2K}$, 
the temperature-dependence of the conductance through a tunnel barrier in a Luttinger liquid\cite{KaneFisher1992},
 $G(T)\propto T^{2(1-K)/K}$, and the temperature scaling of the conductivity of a disordered 
interacting wire\cite{GiamarchiShultz1988GornyiMirlinPolyakov2007}.  There exists by now a rich variety of experimental realizations of LLs with fermionic constituent particles, including  semiconductor, metallic, and polymer nanowires\cite{nanowires}, carbon nanotubes\cite{nanotubes}, edge states of 2D topological 
insulators\cite{top-ins}, and cold-atom systems\cite{cold-atoms}.

 Further, edges of quantum Hall systems\cite{QH-tunneling,QH-MZI,QH-relaxation} give rise to chiral LLs with only one propagation direction.  When two such edges with opposite chirality are coupled by interaction, an artificial ``wire'' emerges \cite{QH-counter,Prokudina}. 
Properties of LL structures are probed in a growing number of sophisticated experiments, in 
particular under strongly non-equilibrium conditions. A quantitative interpretation of experimental findings requires 
the knowledge of the LL parameter $K$ of the studied system by an additional independent measurement of $K$. 

Let us consider a typical experimental setup where a 1D conductor is connected to the outside world by  leads. 
One possibility to access the value of $K$  is to measure the power-law behavior of the TDOS by exploring the tunneling  into the Luttinger liquid.  This requires, however, an introduction of an additional  
probe to the interacting wire and is not simple experimentally. In addition, the tunneling characteristics  may be affected by  the 
interaction of the wire with the environment\cite{GlebFinkelstein}. 
We can  ask if it is possible to infer $K$ considering the LL  as a "black box" (in the spirit of scattering theory of electronic conduction\cite{Buettiker})
and performing  electrical measurements in the leads alone.
 One could naively expect that  the interaction inside the system modifies the conductance of the wire, thus providing a 
direct experimental way to measure $K$. It is not correct, however. It is well known\cite{Safi, Maslov, Ponomarenko, Oreg96} that, due to absence of fermionic backscattering in a clean LL, its DC conductance is given by the interaction-independent value $e^2/h$.  
Moreover, while under generic non-equilibrium conditions the distribution function of the electrons that have passed  the interacting part of the system depends on the interaction strength \cite{Gutman2010_1},  the zero-frequency full counting statistics of the charge transferred through the system\cite{Gutman2010_2} is  insensitive to the interaction. 
On the other hand, the non-equilibrium noise \cite{ac-noise} and the full counting statistics \cite{Gutman2010_2} 
at high frequencies (of the order of or larger than  the inverse flight-time through the system) do depend on the 
interaction strength but they are challenging  to measure experimentally \cite{Kamata2014}.  

In this work we show that the interaction in a LL wire can, however, be probed by low-frequency charge noise measurements provided that the electrons emerging from the LL are mixed (via scattering at an additional quantum point contact, QPC) with electrons coming from an independent  reservoir. A similar approach was proposed recently to probe the (pseudo-)spin-charge separation in systems of co-propagating channels \cite{CopropagatingChannels}.

The structure of the paper is as follows. 
In Sec. \ref{Sec:Setup}  we introduce a device, consisting of 4 sources (SL, SR, S1, S2) and 2 drains D1 and D2
that are connected by two point contacts characterized by transmission and reflection  coefficients $t^2$ and $r^2$ (see Fig 1).
The system is driven out of equilibrium  by an``injection'' of electrons with double-step energy distribution trough the source SR. 
Such a distribution  may be naturally prepared by  means of an additional QPC0 (not shown in Fig. \ref{Fig:setup1}).  The step height $h$ is then given by its transmission coefficient.
In Sec. \ref{Sec:D2} we calculate the shot noise in drain D2 as a function of the Luttinger liquid  parameter K. Section \ref{Sec:D2Gen} is devoted to description of the general formalism while  Sec. \ref{Sec:D2An} summarizes the results in the limits of weak ($|K-1| \ll 1$) and strong ($K\ll 1$)  interaction for  voltage U in SL much larger than the inverse of the flight time $\tau_l$ in the interaction region.  In Sec. \ref{Sec:D2Num} we present few numerical results for generic values of interaction parameter $K$.

Specifically, Figs.  \ref{Fig:noiseLefth} and \ref{Fig:noiseLeftMax} illustrate the central results of the paper. Figure \ref{Fig:noiseLefth} demonstrates the dependence of the noise at zero voltage $V$ (at source $S2$) on the parameter $h$ of the double-step distribution (\ref{Eq:nR}) of incoming right-moving electrons. The noise attains its maximal value when the initial distribution is particle-hole symmetric ($h = 0.5$) and the QPC mixing the electrons from the LL wire
with those from the source S2 has reflection probability $r^2 = t^2 = 0.5$. 
The ratio of this maximal noise to the voltage $U$ in $SL$ is a universal function of the LL parameter.
Fig. \ref{Fig:noiseLeftMax} shows that the maximal current noise at drain D2 at zero frequency and zero voltage in source S2 ($\omega=0$ and $V=0$),   which we denote as $\underset{h,r^2,t^2}{\max}[S_{D2}(\omega=0,V=0)]$,  provides a direct access to the value of the LL parameter K.
Although we do not have a simple analytic expression for the noise in this situation, the curve is universal and our  numerical results can be used to determine K.

Section \ref{Sec:D1} presents the calculation of the noise in drain D1 and Sec. \ref{Sec:attractive} discusses the situation for attractive interactions. We
conclude the manuscript with a summary section, Sec. \ref{Sec:Summary}.

\section{Setup}
\label{Sec:Setup}
\begin{figure}
\includegraphics[width=220pt]{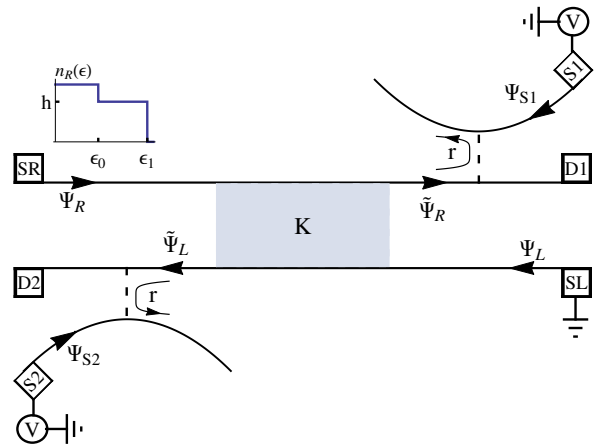}
\caption{\small Setup. Incoming $R$-electrons have a non-equilibrium double-step energy distribution (\ref{Eq:nR}) characterized by the step width $U=\epsilon_1-\epsilon_0$ and height $h$. This distribution may be prepared by  means of a QPC0 (not shown). The parameter $h$ is given in this case by the transmission probability of the QPC0, while the parameter $U$ is the QPC0 voltage. Incoming $L$-electrons, as well as $S1$- and $S2$-electrons are at equilibrium but the distribution of the $S1$ and $S2$ electrons can be tuned with a voltage $V$.}
\label{Fig:setup1}
\end{figure}

A setup that we consider in the present paper (and that is particularly relevant in the context of quantum Hall physics, see, e.g., Ref.~\onlinecite{Prokudina})  is shown in Fig. \ref{Fig:setup1}. 
It includes two counterpropagating electronic (or, more generally, fermionic) modes, right-movers $R$ and left-movers $L$, interacting over a distance $l$ via a short-range interaction characterized by the LL parameter $K$.
The system is driven out of equilibrium  by an``injection'' of incoming $R$-electrons (source SR) with a double-step  energy distribution, 
\begin{equation}
n_R(\epsilon) = (1-h) n_0(\epsilon-\epsilon_0) + h n_0(\epsilon-\epsilon_1) \,,
\label{Eq:nR}
\end{equation} 
Here, $n_0(\epsilon)=\Theta(-\epsilon)$ is the zero-temperature Fermi-Dirac distribution with zero chemical potential 
and  $\epsilon_0=-hU$, $\epsilon_1=(1-h)U$ are the positions of the Fermi edges\cite{Remark:mu}.  The double-step distribution  (\ref{Eq:nR}) may be naturally prepared by  means of a QPC0 (not shown in Fig. \ref{Fig:setup1}). The parameter $h$ is given in this case by the transmission probability of the QPC0, while the parameter $U$ is the QPC0 voltage.  (We set the electron charge $e$ to unity throughout the paper, restoring it in the final expressions only.)
The left-moving mode starts at zero temperature and zero voltage  from the source SL. After traversing the interacting part of the wire, 
right-movers and left-movers are mixed with electrons from sources  S1 and S2  (kept at zero temperature and chemical potential $V$) via scattering at QPCs with transmission (reflection) amplitudes $t$ ($r$). We are interested in the charge noise at drains D1 and D2 
\begin{equation}
 S_{D1/D2}(\omega, V)=\int_\infty^\infty dt\left\langle \left\{\delta I_{D1/D2}(t), \delta I_{D1/D2}(0)\right\}\right\rangle e^{-i\omega t},
\label{Eq:SDef}
 \end{equation}
 where $\delta I_{Di}$ its the fluctuating part of the current operator at the drain $i$, and curly brackets denote the anticommutator. 
The second argument of $S_{D1/D2}$ in Eq. (\ref{Eq:SDef}) emphasizes the dependence of noise on the voltage $V$ applied to the sources S1 and S2. 

\section{Noise at drain D2}
\label{Sec:D2}
\subsection{General formalism}
\label{Sec:D2Gen}
We start with the discussion of the noise at the drain D2, which has a simpler structure. 
To evaluate $S_{D2}$, we express the electronic operator $\Psi_{D2}$ at the drain D2 in terms of the fermionic fields at source S2 and at the output of  the wire (see Fig. \ref{Fig:setup1}),
\begin{equation}
 \Psi_{D2}=t \tilde{\Psi}_L+r\Psi_{S2}.
\end{equation}
Together with the expression for the current operator,  $I_{D2}=u e\Psi^+_{D2}\Psi_{D2}$ (where $u$ is the velocity of the left fermionic mode and $e>0$ is the  absolute value of electron charge), this implies the decomposition of $S_2(\omega, V)$ into 
three contributions, 
$$
S_{D2}=|r|^4 S^{(S2)}_{D2}+|t|^4 S_{D2}^{(SL)}+|r|^2 |t|^2 S_{D2}^{(\rm x)}.
$$ 
The first contribution to the sum represents just the charge noise of zero-temperature non-interacting electrons
coming from the source S2, $S^{(S2)}_{D2}(\omega)\equiv S_0(\omega)=e^2|\omega|/2\pi$. 

The key point for the discussion of the second contribution $S^{(SL)}_{D2}$ (which is also $V$-independent and 
represents the charge noise for fully open QPC) is the observation that it can be expressed solely in terms of the
 electronic density $\tilde{\rho}_L\equiv\tilde{\Psi}^+_L\tilde{\Psi}_L$, $S^{(SL)}_{D2}(t)= u^2\left\langle\left\{\tilde{\rho}_L(t), \tilde{\rho}_L(0)\right\}\right\rangle$. Therefore, the standard bozonization tools can be readily applied. Within the bosonization framework, the effect of the electron-electron interaction in the central part of the wire is to induce the scattering of density fluctuations at  the boundaries of the interaction region. We will assume in the following that the characteristic bosonic momenta (set in our problem by the the voltage $U$ in the distribution function $n_R(\epsilon)$) are small compared to the inverse length of the transition region between the interacting and non-interacting parts of the wire. The bosonic reflection at the wire 
boundary is then characterized\cite{Safi} by the (momentum independent)  amplitude  
\begin{equation}
b=(1-K)/(1+K)
\label{Eq:b}
\end{equation}
and the density $\tilde{\rho}_L$ can be expressed in terms of those at the sources SR  and SL via
\begin{multline}
 \tilde{\rho}_L(t)=(1-b^2)\sum_{n=0}^{\infty}b^{2n}\rho_L(t-(2n+1)\tau_l)\\+b\rho_R(t)-(1-b^2)\sum_{n=0}^{\infty}b^{2n+1}\rho_R(t-(2n+2)\tau_l). 
 \label{Eq:rhoLt}
\end{multline}
Here $\tau_l=Kl/u$ is the time needed to a density perturbation to cross the LL wire. 

Evaluating now the correlation function  of $\tilde{\rho}_L$ with the account of  the fermionic distribution functions in the leads SR and SL, one finds
\begin{equation}
 S^{(SL)}_{D2}(\omega)=\frac{(1-b^2)^2S_0(\omega) }{\left|1-b^2 e^{2i\omega\tau_l}\right|^2}  
+  \frac{2b^2\left[1-\cos(2\omega\tau_l)\right]S_R(\omega)}{\left|1-b^2 e^{2i\omega\tau_l}\right|^2 } ,
\label{Eq:S22}
\end{equation}
where $S_R(\omega)$ is the noise of non-interacting electrons with the double-step distribution~(\ref{Eq:nR}),
$$
S_R(\omega)=\frac{e^2}{2\pi}\left[|\omega|+2h(1-h)(U-|\omega|)\Theta(U-|\omega|)\right] .
$$
Due to  charge conservation and the absence of fermionic backscattering in the wire the zero frequency component of $\tilde{\rho}_L$ in Eq. (\ref{Eq:rhoLt}) 
coincides with that of the incoming density $\rho_L$. As a consequence, $S^{(SL)}_{D2}(\omega)$ is  insensitive to  the interaction strength for low frequencies $\omega \ll 1/\tau_l$ and vanishes at $\omega=0$.

Both contributions $S^{(S2)}_{D2}(\omega)$ and $S^{(SL)}_{D2}(\omega)$ that we have evaluated up to now are thus interaction-independent in the low-frequency regime and vanish at zero frequency.  We turn now to the analysis of the remaining cross-correlation term, $S^{(\rm x)}_{D2}(\omega)$. We will show that it is non-trivial and  interaction-sensitive in the limit of $\omega=0$.  In time domain, this contribution can be presented in the form
\begin{equation}
 S^{(\rm x)}_{D2}(\tau, V)=2 e^2 u^2 \Re\left[G_{S2}^{<}(-\tau)G^{>}_{\tilde{L}}(\tau)+G_{S2}^{>}(-\tau)G^{<}_{\tilde{L}}(\tau)\right].
 \label{Eq:S3Basic}
\end{equation}
Here  $G^{\gtrless}_{S2}(t)=-e^{-i Vt}/2\pi u (t\mp i0)$ stand for the time-domain Keldysh Green functions 
of  $\Psi_{S2}$, while  $G^{\gtrless}_{\tilde{L}}$ are  Green functions of left-moving electrons leaving the interacting wire.  It is not difficult to check that  the zero-frequency noise is directly related to the distribution function of $\Psi_{\tilde{L}}$ electrons, $n_{\tilde{L}}(\epsilon)$, via
\begin{equation}
 \partial_{V}S^{(\rm x)}_{D2}(\omega=0, V)=\frac{e^2}{\pi}[1- 2 n_{\tilde{L}}(V)].
 \label{Eq:NoiseDistribution}
\end{equation}

The non-equilibrium bosonization technique for evaluation  of correlation functions in a far-from-equilibrium LL in the framework of Keldysh formalism was developed in Refs. \onlinecite{Gutman2010_1, Protopopov2011, Protopopov2013}. It was shown, that arbitrary (including many-particle) electronic correlation function can be expressed as a product of Fredholm determinants, $G\sim \Delta_R[\delta_R(t)]\Delta_L[\delta_L(t)]$, 
 having the form
\begin{equation}
 \Delta_\eta[\delta_\eta(t)] =\det\left[1+\left(e^{i\delta_\eta(t)}-1\right)n_\eta(\epsilon)\right].
 \label{Eq:DetGeneral}
\end{equation}
Here, $n_{R(L)}(\epsilon)$ denotes the distribution function of electrons injected into the LL from the right (left) lead, while the phases $\delta_\eta(t)$ encode the information on the  correlation function of interest and on the interaction strength in the wire. The time $t$ and energy $\epsilon$ in Eq.(\ref{Eq:DetGeneral}) are understood as canonically conjugate variables, $\epsilon = i\hbar (\partial/\partial t)$.
Fredholm determinants of a type similar to Eq.~(\ref{Eq:DetGeneral}) arise also in the theory of full counting statistics \cite{FCS}, non-equilibrium Fermi-edge singularity \cite{fermi-edge}, chiral 1D systems (including quantum Hall Mach-Zehnder interferometry) \cite{QH-theory}, and spectral functions of nonlinear Luttinger liquids \cite{nonlinear-LL}. 

\begin{figure}
\includegraphics[width=220pt]{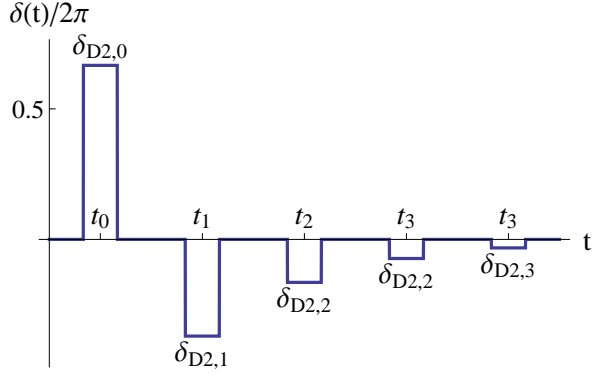}
\caption{\small Phase $\delta(t)$ determining the charge noise at the drain D2, see Eqs. (\ref{Eq:DetNormalized}), (\ref{Eq:S2Final}). The phases $\delta_{D2, n}$ are given by Eq. (\ref{Eq:deltaL}).  We have assumed $K=0.2$ to generate the plot. }
\label{Fig:leftpulses}
\end{figure} 

Evaluating  the single-particle correlation functions of the electrons leaving the LL [which are involved in Eq.(\ref{Eq:S3Basic})] by means of this technique and taking into account the zero-temperature electronic distribution in the source $S2$ (see Fig. \ref{Fig:setup1}), we get
\begin{equation}
G^{\gtrless}_{\tilde{L}}(\tau)=-\frac{1}{2\pi u(\tau\mp i0)}\bar{\Delta}(\tau)
\label{Eq:GDet}
\end{equation}
Here $\bar{\Delta}(\tau)$ stands for the  Fredholm determinant (\ref{Eq:DetGeneral}) normalized to its zero-temperature value
\begin{equation}
\bar{\Delta}(\tau)=\frac{\det\left[1+\left(e^{i\delta(t)}-1\right)n_R(\epsilon)\right]}{\det\left[1+\left(e^{i\delta(t)}-1\right)n_0(\epsilon)\right]} \,,
\label{Eq:DetNormalized}
\end{equation}
with the phase $\delta(t)$ being a sequence of rectangular-shaped pulses of duration $\tau$  centered 
at $t_n= 2 n \tau_l$, $n=0,\, 1,\, \ldots$ (see Fig. \ref{Fig:leftpulses}):
\begin{eqnarray}
 \delta(t)&=&\sum_{n=0}^{\infty}\delta_{D2, n} \Theta(t-t_n-\tau/2)\Theta(-t+t_n+\tau/2); \ \ 
 \label{Eq:deltaLGeneral}\\
 \delta_{D2, n}&= &2\pi \left\{ \begin{matrix} b \qquad &\text{for} &n=0& \\ -(1-b^2)b^{2n-1} \qquad &\text{for} &n>0& \,. \end{matrix} \right.
 \label{Eq:deltaL}
\end{eqnarray}
Substitution of Eq. (\ref{Eq:GDet}) into Eq. (\ref{Eq:S3Basic}) yields
\begin{multline}
 S^{(\rm x)}_{D2}(\omega, V)=2S_0(\omega)\\ -\frac{e^2}{\pi^2}\int  d\tau ~\frac{\Re \left(\bar{\Delta}(\tau) e^{iV\tau}-1 \right)}{\tau^2}\cos(\omega\tau). 
 \label{Eq:S23}
\end{multline}
In the low-frequency limit, we thus obtain for the noise of electrons at the drain D2:
\begin{equation}
 S_{D2}(\omega=0, V)= -\frac{e^2 |t|^2 |r|^2}{\pi^2}\int  d\tau ~\frac{\Re \left(\bar{\Delta}(\tau) e^{i e V\tau}-1 \right)}{\tau^2}.
\label{Eq:S2Final}
 \end{equation}

We proceed now to the analysis of the result (\ref{Eq:S2Final}).
The $\tau$-dependence of the normalized determinant $\bar{\Delta}(\tau)$ is characterized by two distinct time scales. First, there is the  scale $\tau_l=Kl/u$ set by the length of the 
interacting wire which determines the time interval between successive  pulses in Eq. (\ref{Eq:deltaLGeneral}).
Under the assumption $\tau\ll \tau_l$ we can approximate the determinant $\bar{\Delta}(\tau)$ by a product of Toeplitz determinants corresponding to the individual rectangular-shaped pulses in Eq. (\ref{Eq:deltaLGeneral}),
\begin{equation}
 \bar{\Delta}(\tau)=\prod_{n=0}^\infty \bar{\Delta}(\tau, \delta_n).
 \label{Eq:DeltaProduct}
\end{equation}
The second scale controls the long-time behavior of Eq.~(\ref{Eq:DeltaProduct}). 
According to the asymptotic theory of Toeplitz determinants reviewed in Refs. \onlinecite{Gutman2011, Protopopov2013}
the long-time asymptotics of $\bar{\Delta}(\tau)$ is controlled by the exponential decay, $\bar{\Delta}(\tau)\propto e^{-\tau/2\tau_\phi}$, with a non-equilibrium dephasing rate given by
\begin{equation}
 \frac{1}{\tau_\phi}=-\frac{U}{\pi}\sum_{n=0}^\infty \Re\ln \left[1+(e^{-i\delta_n}-1)h\right].
 \label{Eq:dephasing}
\end{equation}
For the generic parameters, the dephasing rate (\ref{Eq:dephasing}) is of the order of $U$, thus setting a characteristic time scale $1/U$. 
Therefore, the integral (\ref{Eq:S2Final}) converges at times $\tau$ set\cite{Remark:LongTime} by  $1/\rm{max}\{U,V\}$.
We thus conclude that the approximation Eq. (\ref{Eq:DeltaProduct}) for the determinant entering Eq.~(\ref{Eq:S2Final}) can be safely applied provided that at least one of voltages $U$ and $V$ is large compared to the inverse flight time $1/\tau_l$.

\subsection{Asymptotic behavior}
\label{Sec:D2An}
Equations (\ref{Eq:S2Final}) and (\ref{Eq:DeltaProduct}) render the current noise $S_{2D}(\omega=0, V)$ 
amiable to straightforward numerical evaluation (see, e.g., Ref. \onlinecite{Protopopov2012} for a detailed account of numerical procedure).
Two limiting cases  can also be studied analytically. First, in the  weak-interaction regime, $|K-1|\ll 1$, all the phases $\delta_{D2, n}$ are small, enabling perturbative evaluation of the determinant $\bar{\Delta}(\tau)$. The result reads
\begin{eqnarray}
\ln \bar{\Delta}(\tau)&=&-\alpha_1 \int_0^U d\omega \frac{\sin^2\omega \tau/2}{\omega^2}(U-\omega),
\label{Eq:DetK1D2}
\\
\alpha_1&=& 2(K-1)^2 h(1-h).
\label{Eq:alphaK1D2}
\end{eqnarray}
with the asymptotic behavior
\begin{equation}
  \ln \bar{\Delta}(\tau)= -\frac{\pi\alpha_1 U\tau}{4}+\frac{\alpha_1}{2} \ln U\tau, \qquad  U\tau\gg 1.
  \label{Eq:DetK1D2Asymptotic}
\end{equation}
Examining the integral (\ref{Eq:S2Final}), we see that at $V\gg \alpha_1 U$ we can further expand $\bar{\Delta}(\tau)=\exp\left(\ln \bar{\Delta}(\tau) \right)$ in powers of $\alpha_1$.
On the other hand, for $V\ll \alpha_1 U$ the integral (\ref{Eq:S2Final}) is dominated by  long times where the asymptotics (\ref{Eq:DetK1D2Asymptotic}) applies.
We thus obtain the following result for the noise in the case of a weak interaction:
\begin{equation}
 S_{D2}(\omega=0, V)\underset{|K-1|\ll 1}{=}
 \frac{e^2 |t|^2|r|^2 U}{\pi}f(\alpha_1,V/U),
 \label{Eq:S2K1}
\end{equation}
with the small parameter $\alpha_1$  given by Eq. (\ref{Eq:alphaK1D2}) and with a dimensionless function $f$ given by
\begin{multline}
 f(\alpha, x)=\\
 =\left\{
  \begin{array}{cc}
|x|, & x> 1\\
  |x|+\alpha\left[|x|-1-\frac{1+|x|}{2}\ln |x|\right], &  \alpha\ll|x|<1 \\
  -\frac\alpha2+\frac2\pi x \arctan\frac{4x}{\pi\alpha}-\frac\alpha4\ln\left(x^2+\frac{\pi^2\alpha^2}{16}\right), & |x|\lesssim \alpha.
  \end{array}
  \right.
 \label{Eq:f}
\end{multline}

The second limit that can be treated fully analytically is that of a strong repulsive interaction, $K\ll 1$. In this situation, the bosonic reflection coefficient is close to  $1$. As a consequence, all but the first phase differences 
$\delta_{D2, n>0}$ are small, while $\delta_{D2, 0}$ is close to $2\pi$. The determinant $\bar{\Delta}(\tau, \delta)$ at the phase $\delta=2\pi$ corresponds 
to free fermions and can be computed exactly via the refermionization procedure,
\begin{equation}
 \bar{\Delta}(\tau, 2\pi)=(1-h)e^{-i\epsilon_0\tau}+he^{-i\epsilon_1\tau}.
\label{Eq:Det2Pi}
 \end{equation}
Treating all but the first factors in Eq. (\ref{Eq:DeltaProduct}) perturbatively [cf. Eqs. (\ref{Eq:DetK1D2}) and~(\ref{Eq:alphaK1D2})] and also taking into account the perturbative correction 
to the zeroth-order approximation (\ref{Eq:Det2Pi}) for the determinant $\bar{\Delta}(\tau, \delta_{D2, 0})$, we arrive at
\begin{multline}
\bar{\Delta}(\tau)=\left[\bar{\Delta}(\tau, 2\pi)-\frac{\alpha_0}{2}\left(e^{-i\epsilon_0\tau}g(U\tau)+e^{-i\epsilon_1\tau}g(-U\tau)\right)\right]\\
\times
\exp\left[-\alpha_0 \int_0^U d\omega \frac{\sin^2\omega \tau/2}{\omega^2}(U-\omega)\right]
\label{DetSmallKD2}
\end{multline}
with 
\begin{eqnarray}
\alpha_0&=&8K h (1-h),
\label{Eq:alphaSmallKD2}
\\
g(x)&=&\gamma- \Ci(x) + \ln x + i \Si(x).
\end{eqnarray}
Here $\gamma$ is the Euler constant, while $\Ci(x)$ and $\Si(x)$ stand for the integral cosine and sine functions. 

To zeroth order in the small parameter $\alpha_0$, Eq. (\ref{Eq:alphaSmallKD2}), we get  
\begin{equation}
S_{D2}(\omega=0, V)\underset{K= 0}{=}\frac{e^2|t|^2|r|^2}{\pi}\left[(1-h)|V-\epsilon_0|+h|V-\epsilon_1|\right].
\label{Eq:SD2K0}
\end{equation}
An analysis analogous to the one that leaded to Eqs. (\ref{Eq:S2K1}) and (\ref{Eq:f}) allows us to establish also the first correction~to~Eq.~(\ref{Eq:SD2K0}), yielding 
\begin{multline}
 S_{D2}(\omega=0, V)\underset{K\rightarrow 0}{=}
 \frac{e^2 U|t|^2|r|^2}{\pi}\\
 \times
 \left\{
  (1-h)f\left(\alpha_0, \frac{V-\epsilon_0}{U} \right)+
   h f\left(\alpha_0, \frac{V-\epsilon_1}{U} \right)\right.\\\left.
    +\alpha_0\left[ p\left(\frac{V-\epsilon_0}{U}\right)+ p\left(\frac{\epsilon_1-V}{U}\right)\right]\Theta(V>\epsilon_0)\Theta(V<\epsilon_1)
  \right\}.
  \label{Eq:S2K0}
\end{multline}
with  $ p(x)\equiv x\ln x$ and $\alpha_0$ given by Eq. (\ref{Eq:alphaSmallKD2}).

\begin{figure}
\includegraphics[width=220pt]{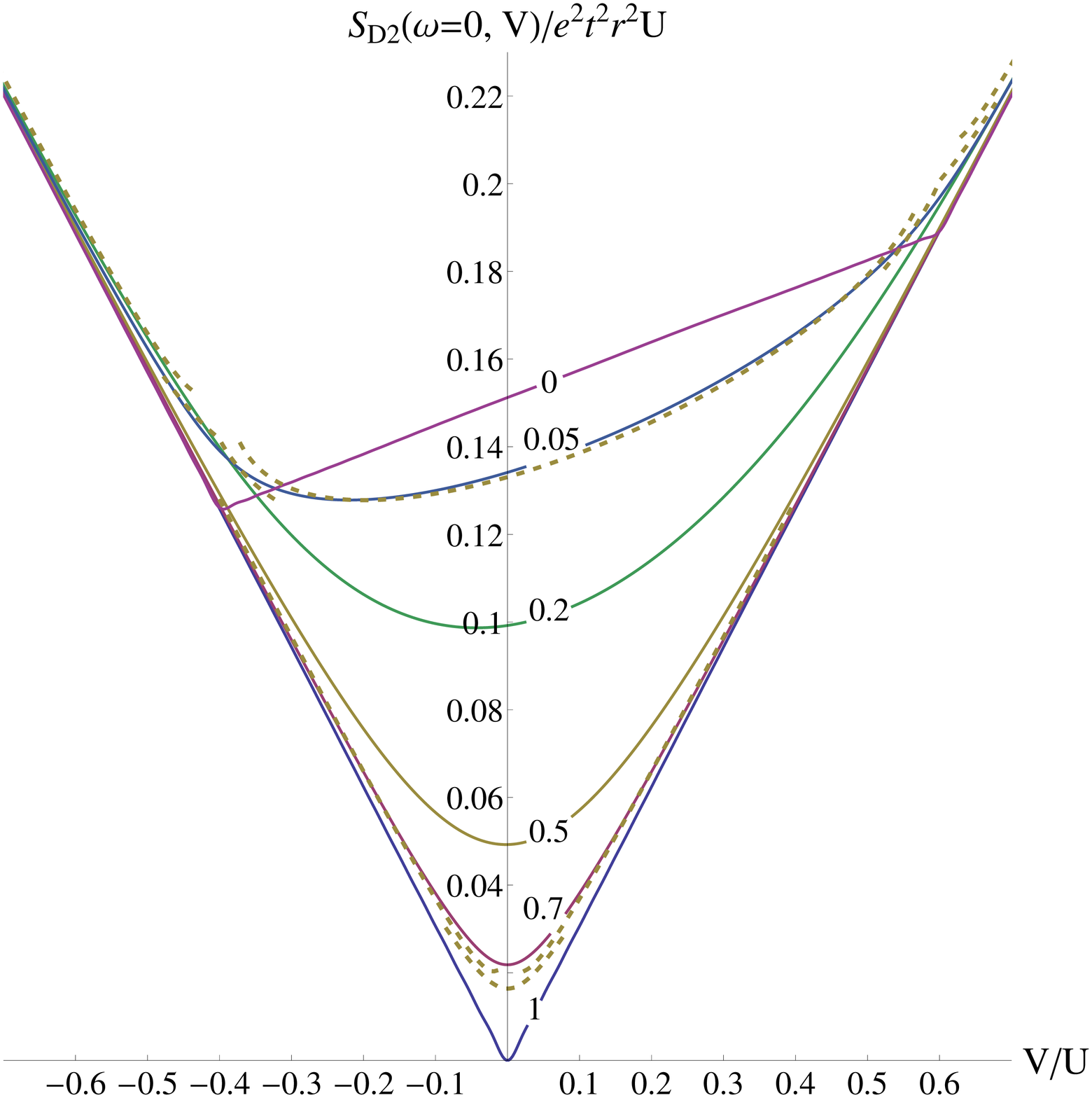}
\caption{\small Zero-frequency noise at the drain D2,  $S_{D2}(\omega=0, V)$, 
 given by Eq.~(\ref{Eq:S2Final}),
as a function of the voltage ratio $V/U$  for different values of the interaction strength. 
The parameter $h$ is chosen to be $0.4$. The curves are labeled according to the LL parameter $K$ characterizing 
the interaction strength. 
Dashed curves represent appropriate asymptotic expressions [Eqs. (\ref{Eq:S2K1}), (\ref{Eq:f}) for a 
relatively weak interaction, $K=0.7$, and Eqs. (\ref{Eq:S2K0}), (\ref{Eq:f}) for a strong interaction, $K=0.05$].  
}
\label{Fig:noiseLeft}
\end{figure}

\begin{figure}
\includegraphics[width=220pt]{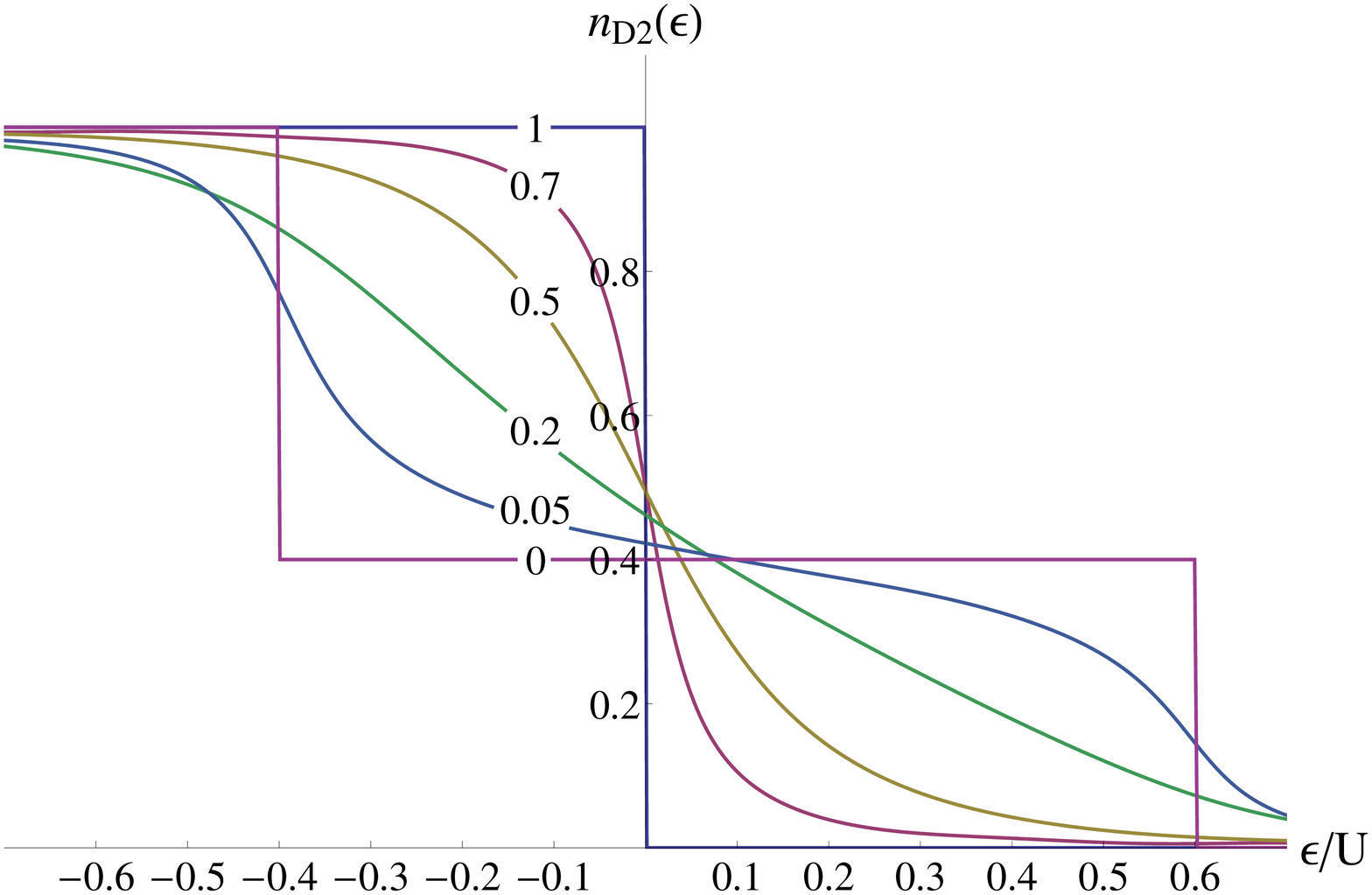}
\caption{\small Distribution of electrons over energies  at the drain D2,  $n_{D2}(\epsilon)$,  for different values of the interaction strength. 
The parameter $h$ is chosen to be $0.4$. The curves are labeled according to the LL parameter $K$ characterizing the interaction strength.
}
\label{Fig:nLeft}
\end{figure}

\begin{figure}
\includegraphics[width=220pt]{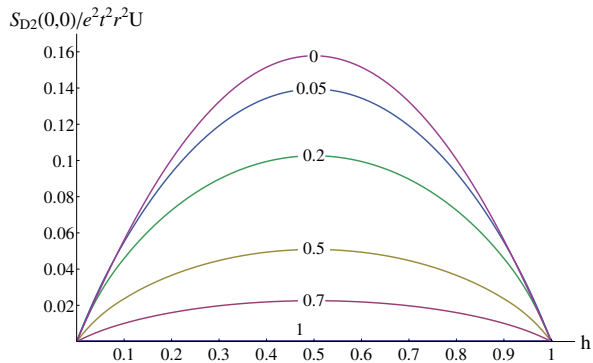}
\caption{\small Noise at the drain D2 at zero frequency and zero voltage $V$,  $S_{D2}(\omega=0, V=0)$, 
as a function of the step height $h$ for different values of the interaction strength.
 The curves are labeled according to the LL parameter $K$ characterizing the interaction strength. } 
\label{Fig:noiseLefth}
\end{figure} 

\begin{figure}
\includegraphics[width=220pt]{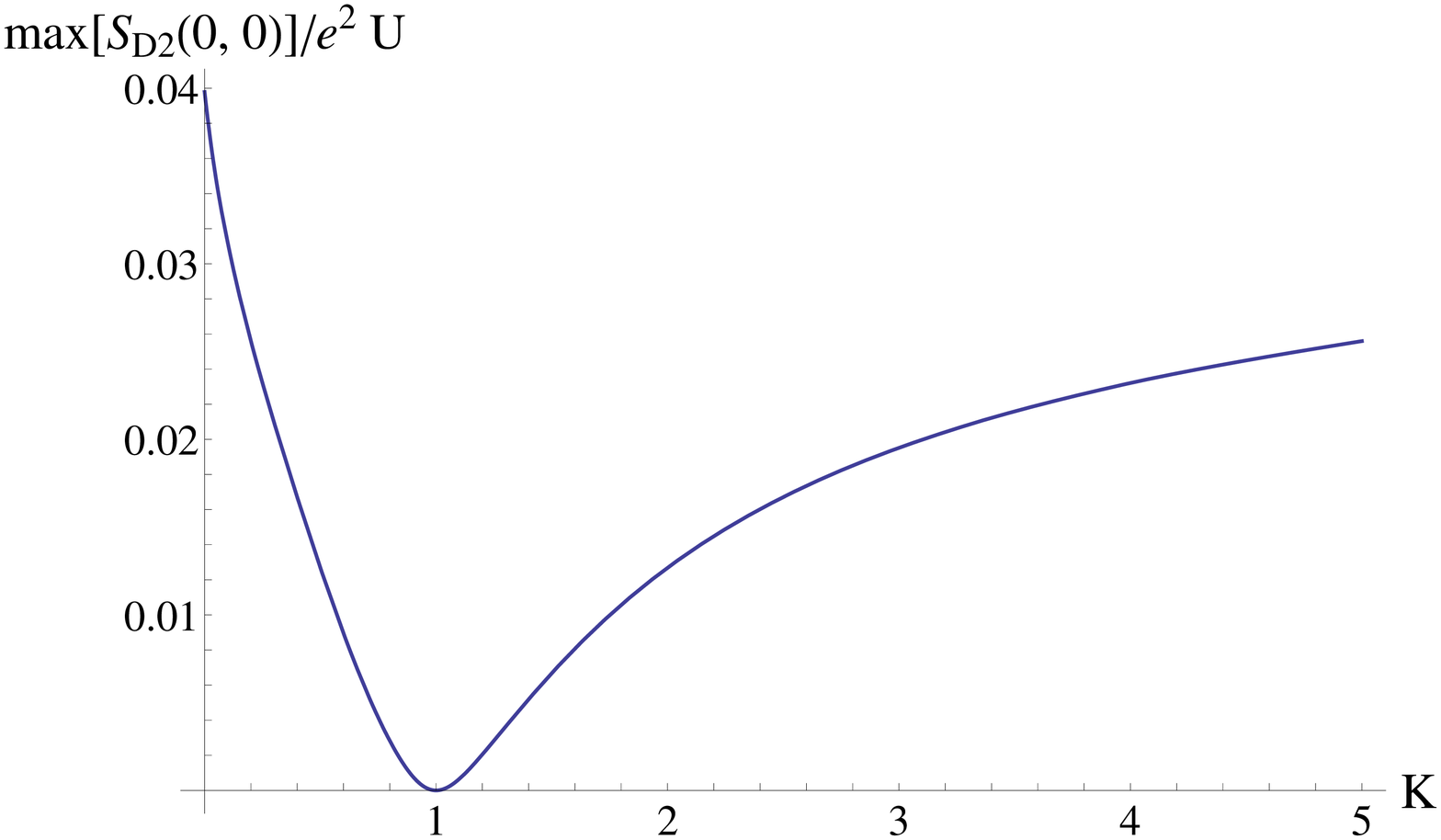}
\caption{\small The maximal noise in drain D2 at zero frequency and zero voltage $V$,  $\underset{h, r^2, t^2}{\max}S_{D2}(\omega=0, V=0)$, 
as a function of the Luttinger liquid parameter  $K$. The noise attains maximum when the initial distribution is particle-hole symmetric ($h=0.5$) and the QPC mixing the electrons from the LL wire with those from the source 
S2 has reflection probability $r^2=t^2=0.5$, see Sec. \ref{Sec:D2Num}.
The maximal noise  equals $1/8\pi$ at $K=0$ and approaches the same value asymptotically as $K\rightarrow \infty$ (see Sec. \ref{Sec:attractive} 
for the discussion of the case of attractive interaction $K>1$).   } 
\label{Fig:noiseLeftMax}
\end{figure} 

\subsection{Numerical evaluation}
\label{Sec:D2Num}
We turn now to the case of a generic interaction strength, when the determinant in Eqs. (\ref{Eq:S2Final}) and (\ref{Eq:DeltaProduct}) is computed numerically.
Figures \ref{Fig:noiseLeft} and  \ref{Fig:nLeft} present the resulting evolution of the noise $S_{D2}(\omega=0, V)$ 
 [as given by Eq.~(\ref{Eq:S2Final})]
and of the distribution function 
$n_{D2}(\epsilon)$ upon variation of the LL parameter $K$ characterizing the interaction strength. We have set $h=0.4$ to generate the plots. 
The dashed curves provide the comparison to the appropriate asymptotic expressions [(\ref{Eq:S2K1}) for $K=0.7$ and (\ref{Eq:S2K0}) for $K=0.05$].
Upon increase of the interaction strength, the distribution function of outgoing electrons $n_{D2}$ evolves from the zero-temperature distribution of the incoming left-moving electrons towards the double-step distribution of the electrons at source SR. It is interesting to note that the characteristic width of the distribution function is non-monotonous as a function of the interaction strength, 
which is related to the non-monotonous dependence of the dephasing rate (\ref{Eq:dephasing}) on the Luttinger parameter $K$. 

Figure \ref{Fig:noiseLefth} demonstrates the dependence of the noise at zero voltage $V$ on  the parameter $h$ of the double-step distribution (\ref{Eq:nR}) of incoming right-moving electrons. The step height $h$ can be experimentally varied by changing the transmission of the QPC0 creating the initial distribution.
The noise attains the maximal value when the initial distribution is particle-hole symmetric ($h=0.5$) and the QPC mixing the electrons from the LL wire with those from the source 
S2 has reflection probability $r^2=t^2=0.5$. The ration of this maximal noise to the voltage $U$ is a universal function of the LL parameter. 
Figures  \ref{Fig:noiseLeft}, \ref{Fig:noiseLefth} and \ref{Fig:noiseLeftMax} show that 
the current noise $S_{D2}(\omega=0)$ provides a direct access to the value of the LL parameter $K$.

\section{Noise at  drain D1}
\label{Sec:D1}
\begin{figure}
\includegraphics[width=220pt]{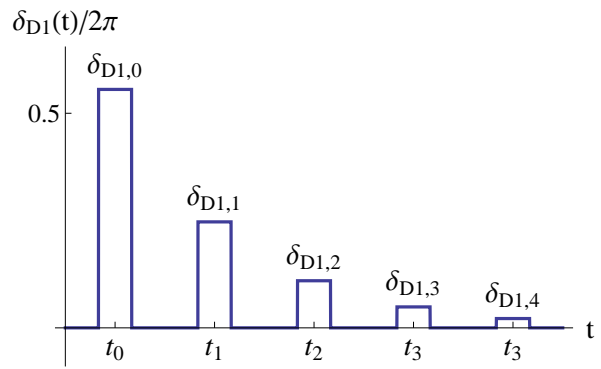}
\caption{\small Phase $\delta(t)$ determining the charge noise at the drain D1, see Eqs. (\ref{Eq:DetNormalized}), (\ref{Eq:S1Final}). The phases $\delta_{D1, n}$ are given by Eq. (\ref{Eq:deltaR}).  We have assumed $K=0.2$ to generate the plot.} 
\label{Fig:rightpulses}
\end{figure}
\begin{figure}
\includegraphics[width=220pt]{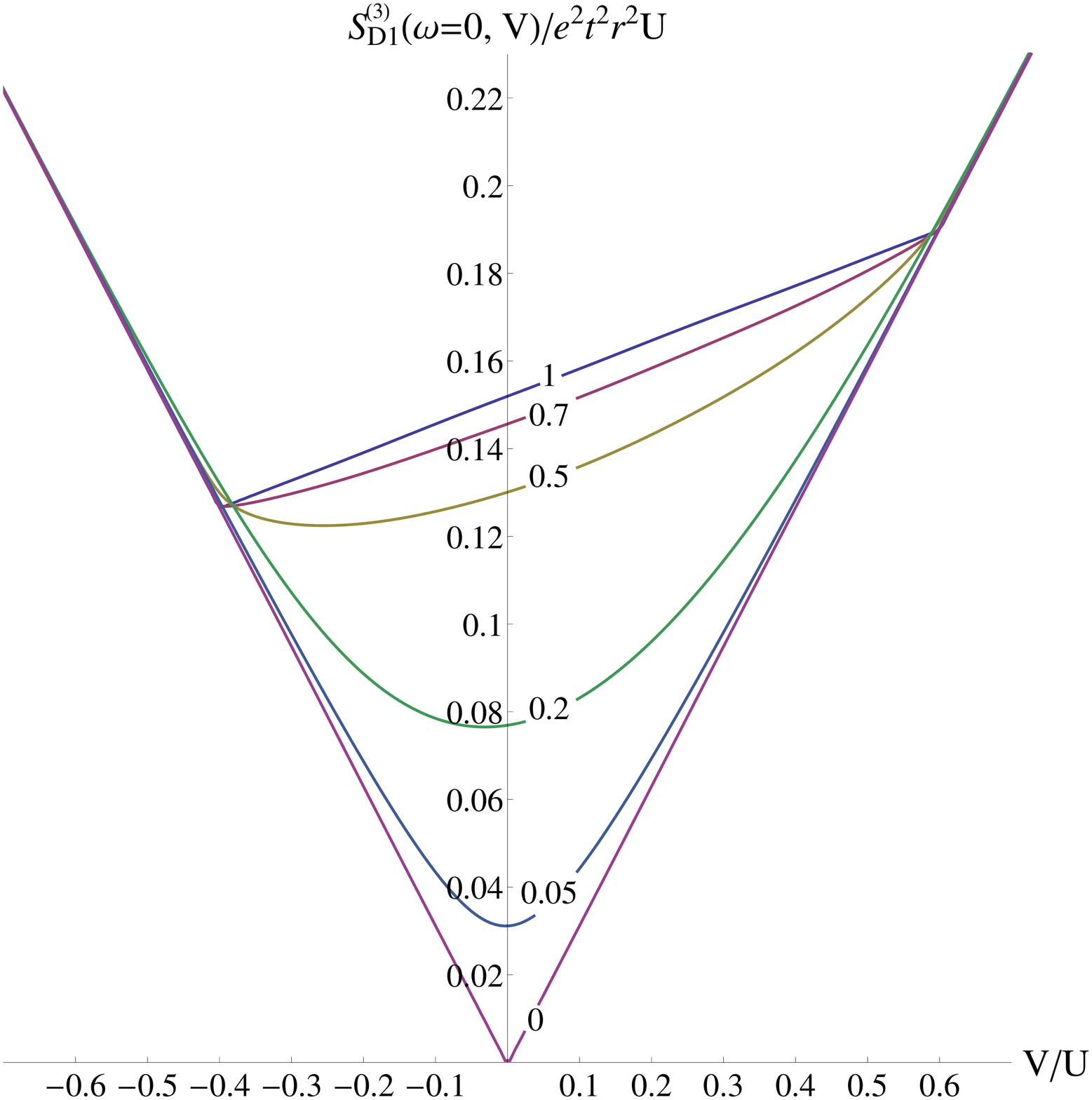}
\caption{\small Zero-frequency noise at the drain D1,  $S_{D1}^{(\rm x)}(\omega=0, V)$, as a function of the voltage $V$ for different values of the interaction strength. 
The parameter $h$ is chosen to be $0.4$. The curves are labeled according to the LL parameter $K$ characterizing the interaction strength. }
\label{Fig:noiseRight}
\end{figure}
\begin{figure}
\includegraphics[width=220pt]{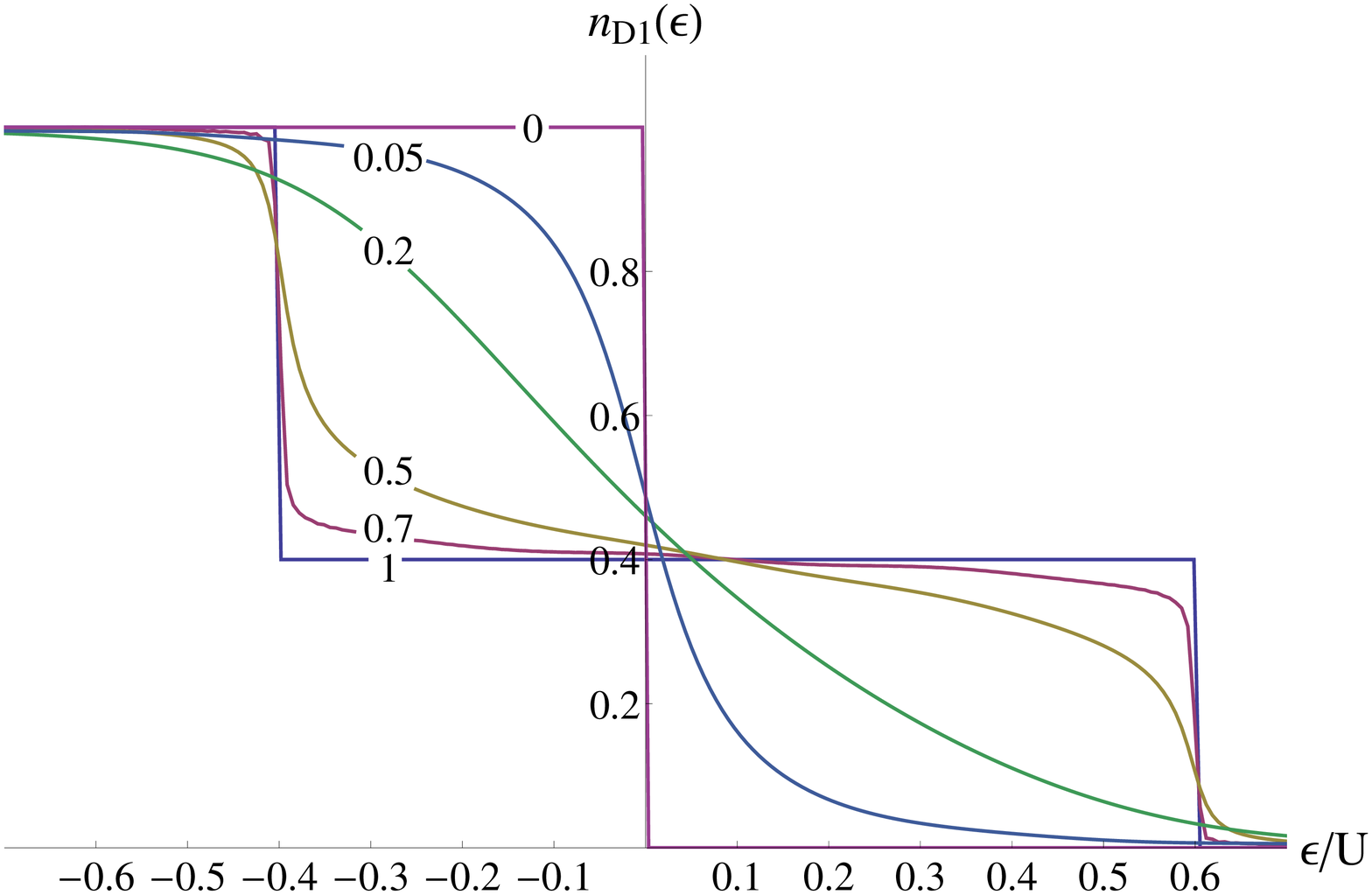}
\caption{\small Distribution of the electrons over energies  at the drain D1,  $n_{D1}(\epsilon)$,  for different values of the interaction strength. 
The parameter $h$ is chosen to be $0.4$. The curves are labeled according to the LL parameter $K$ characterizing the interaction strength.}
\label{Fig:nRight}
\end{figure}
Let us now consider briefly the noise at the drain  D1. In complete analogy to $S_{D2}$, it  can be written as
\begin{equation}
S_{D1}(\omega, V)=|r|^4 S^{(S1)}_{D1}+|t|^4 S_{D1}^{(SR)}+|r|^2 |t|^2 S_{D1}^{(\rm x)}.
\label{d1-noise}
\end{equation}
In contrast to the case of the noise at D2 the term  $S_{D1}^{(SR)}(\omega)$ does not vanish at zero frequency,
$S_{D1}^{(SR)}(\omega=0)=e^2 h(1-h) U/\pi$.   However, in full similarity with the D2 noise, only the last contribution 
in Eq.~(\ref{d1-noise}) is
sensitive to the voltage $V$ at the QPC.
Concentrating on the $V$-dependence of the noise, we thus observe that Eq. (\ref{Eq:S2Final}) applies,
\begin{equation}
 S_{D1}^{(\rm x)}(\omega=0, V)= -\frac{e^2}{\pi^2}\int  d\tau ~\frac{\Re \left(\bar{\Delta}(\tau) e^{i e V\tau}-1 \right)}{\tau^2},
\label{Eq:S1Final}
 \end{equation}
where the phase $\delta(t)$ is now given by (see Fig. \ref{Fig:rightpulses})
\begin{eqnarray}
 \delta(t)&=&\sum_{n=0}^{\infty}\delta_{D1, n} \Theta(t-t_n-\tau/2)\Theta(-t+t_n+\tau/2),\quad
 \\
 \delta_{D1, n}&= & 2\pi (1-b^2) b^{2n}.
 \label{Eq:deltaR}
\end{eqnarray}
Figures \ref{Fig:noiseRight} and \ref{Fig:nRight} show results of a numerical evaluation of the noise $S^{(\rm x)}_{D1}$ and the distribution of the electrons at drain D1. 
The perturbative analysis of the D1 noise in the weak- and strong-interaction limits can be obtained by a straightforward generalization of the corresponding results for the noise D2 presented above. We reiterate that although the D1 noise contains an additional contribution $S^{(SR)}_{D1}$, this contribution is independent on the voltage $V$. Therefore, the $V$ dependence of the D1 noise can also be used (along with the D2 noise) for extracting the LL interaction parameter $K$.

\section{Attractive interaction}
\label{Sec:attractive}
So far [and in particular in the discussion of the strong interaction limit of the model, Eq. (\ref{Eq:S2K0})] we were mostly 
concentrating on the situation of repulsive interaction in the system, $K<1$. The extension of our results to the case 
of attractive 
interaction, $K>1$ (particularly relevant in the context of superconducting wires\cite{Giamarchi}, 
cold atoms\cite{cold-atoms} and fractional quantum Hall systems\cite{QH-counter}),  
is straightforward due to the symmetry of the bosonic reflection
 amplitude $b$, Eq. (\ref{Eq:b}),
\begin{equation}
 b(1/K)=-b(K).
\label{Eq:bSymmetry}
\end{equation}
Equation (\ref{Eq:bSymmetry}) implies  that the charge noise $S_{D1}(\omega, V)$ at drain D1 is invariant with
 respect to  the transformation  $K\rightarrow 1/K$, while the noise  and the distribution function at D2 obey
\begin{eqnarray}
 S_{D2}(\omega, V, 1/K)&=&S_{D2}(\omega, -V, K),\\
\tilde{n}_L(\epsilon, 1/K)&=&1-\tilde{n}_L(-\epsilon, K).
\end{eqnarray}
In particular, in the  limit of infinitely strong attractive interaction, $K=\infty$, 
\begin{equation}
\tilde{n}_L(\epsilon)=1-n_R(-\epsilon)
\label{Eq:nLTildeInf}
\end{equation}
due to Andreev reflection of electrons on the boundary of interaction region.

  
\section{Summary}
\label{Sec:Summary}

To summarize, we have studied the low-frequency noise of interacting electrons in a 1D 
structure with counterpropagating modes (quantum wire), assuming a single channel in each direction. 
Experimental realizations of such structures include also artificial quantum wires formed by counter propagating quantum
 Hall channels coupled by the interaction.
The system is driven out of equilibrium by a QPC0 with an applied voltage, which induces a double-step
 energy distribution of incoming electrons on one side of the device. A second QPC serves to explore the statistics of
 outgoing electrons. We evaluate the dependence of the zero-frequency noise on $K$ and on parameters of both QPCs
 (transparencies and voltages). 

 Our general result, Eq. (\ref{Eq:S2Final}), expresses the noise in the drain D2 in terms of a Fredholm determinant $\bar{\Delta}(\tau)$. In the limits of weak and 
 strong interaction analytical asymptotic (\ref{Eq:S2K1})  and (\ref{Eq:S2K0}) have been obtained. For a generic interaction strength, the noise can be readily evaluated numerically, as shown in Figs. \ref{Fig:noiseLeft}, \ref{Fig:noiseLefth}, \ref{Fig:noiseLeftMax}. Similar results hold for the noise in the drain D1.
 Our findings demonstrate that measurement of a low-frequency noise in such a
 setup allows one to extract the information about the Luttinger liquid constant $K$ which is the key parameter 
characterizing an interacting 1D system.

Upon completion of this work we have learned about a related activity on the noise in systems of co-propagating channels\cite{KovrizhinUnp}.

\begin{acknowledgments}

We acknowledge financial support by DFG Priority Program 1666, by German-Israeli Foundation, and by the EU Network Grant InterNoM. The research of A.D.M. was supported by the Russian Science Foundation (project 14-22-00281). Y.O. acknowledges the support of the Israeli Science Foundation (ISF), the Minerva foundation, and
the European Research Council under the European Community's Seventh Framework Program (FP7/2007-
2013)/ERC Grant agreement No. 340210.

\end{acknowledgments}



\begin{thebibliography}{99}
\bibitem{Remark:mu} For the sake of clarity we assume throughout the paper  that the effective chemical potential defining the electronic 
density in the source SR equals zero,  $(1-h)\epsilon_0 +h \epsilon_1=0$. Upto trivial energy shifts our results are, however, 
independent of this assumption.
\bibitem{Remark:LongTime} There are exceptional situations when the normalized determinants $\bar{\Delta}(t)$ 
exhibits oscillations at frequencies $\epsilon_0$ and $\epsilon_1$ damped only weakly ($ U\tau_\phi\gg 1$ ). 
This is the case for example in the strong-interaction limit $K\ll 1$, discussed in the main text. In such a situation the 
incoherent approximation (\ref{Eq:DeltaProduct}) applies not to close to the Fermi edges $\epsilon_{0, 1}$, 
$|V-\epsilon_{0,1}|\gg 1/\tau_l$.
\end{thebibliography}
\end{document}